\documentclass{llncs}
\usepackage{makeidx}  
\usepackage{graphicx}
\usepackage{subcaption}

\begin{document}
\frontmatter          
\pagestyle{headings}  
\title{Radiation Monitoring with Diamond Sensors for the Belle-II Vertex Detector}
%
%
\author{Chiara La Licata on behalf of the BEAST II collaboration}
\institute{INFN and University of Trieste, Italy\\
\email{chiara.lalicata@ts.infn.it}
}

\maketitle              

\begin{abstract}
The Belle II detector is currently under construction at the SuperKEKB electron-positron high-luminosity collider, that will provide an instantaneous luminosity 40 times higher than that of KEKB. Therefore the Belle-II VerteX Detector (VXD) will operate in a very harsh environment. A radiation monitoring and beam abort system is needed to safely operate the VXD detector in these conditions. 
This system is based on 20 single crystal CVD diamond sensors placed in 20 key positions in the vicinity of the VXD and interaction region.
In this contribution we describe the system design and we present the procedures followed for the characterisation and calibration of the diamond sensors. We discuss also the performance of the prototype system during the first SuperKEKB commissioning phase in February-June 2016.
\end{abstract}
\section{Introduction}
The Belle II detector~\cite{ref:belle2}, at its final construction and commissioning stage, is currently under construction at the KEK laboratory in Tsukuba (Japan). 
Continuing the investigation started by the Belle experiment~\cite{ref:belle}, the aim of Belle II is to probe fundamental physics issues including studies of the CP violation in the flavour sector and rare B decays, searching for deviation from the Standard Model. In order to perform high precision measurements all components of the detector have been redesigned, in particular the VXD, with its first layers very close to the interaction region. High statistics will be achieved with the SuperKEKB collider upgrade whose designed instantaneous luminosity will be 40 times higher than that of KEKB~\cite{ref:kekb}. A drawback of the high luminosity is the increase of radiation level especially in the Vertex Detector volume; the radiation monitoring system~\cite{ref:monitoring} assumes the important role of detecting potentially damaging beam losses.

\section{The Radiation Monitoring System}
\begin{figure}[!h]
\centering
\includegraphics[width=3.2in]{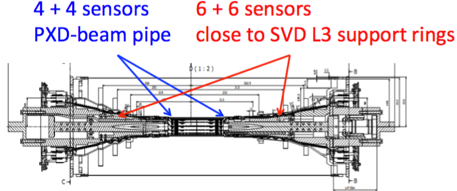}
\caption{Locations of the 20 single crystal diamond sensors of the radiation monitoring system for the inner vertex detector~\cite{ref:monitoring}.}
\label{fig:monitoringSystemScheme}
\end{figure}

The radiation monitoring system will provide an accurate measurement of instantaneous and integrated dose, detecting both sudden large increases of the beam-induced background that could damage the detector, and less important increases that could, if persisting, bring to unacceptable integrated dose. 
The system will be composed of 20 radiation-hard single-crystal Chemical Vapour Deposition (sCVD) diamond sensors, placed close to the nominal interaction point inside the Vertex Detector volume (fig.~\ref{fig:monitoringSystemScheme}).
Properties like the large band-gap (5.5 eV) and high displacement energy (42 eV) make diamond an ideal radiation hard material for particle detectors. Moreover intrinsic diamond sensors do not need darkness, pn-doping or cooling for high-performance operation.

\vspace{-0.5cm}

\section{Calibrations of Diamond Sensors}
The growing process of the diamond crystal can introduce impurities in the bulk that can cause a non uniform response of the diamond sensor. Other effects in the response can arise because of the properties of the interface between the bulk and the electrodes. Additional charge can be injected into the bulk by the electrodes under conditions of large space-charge generated by ionising particles (``photoconductive gain'').
For these reasons, the sensors need to be completely calibrated before the installation inside the Belle II detector: checking the stability in time, the uniformity of the electric field inside the crystal, measuring the charge collection efficiency and evaluating the conversion factor between the measured current and the dose rate. 

\begin{figure}[htbp]
    \begin{minipage}{0.45\textwidth}
        \includegraphics[width=2.3in]{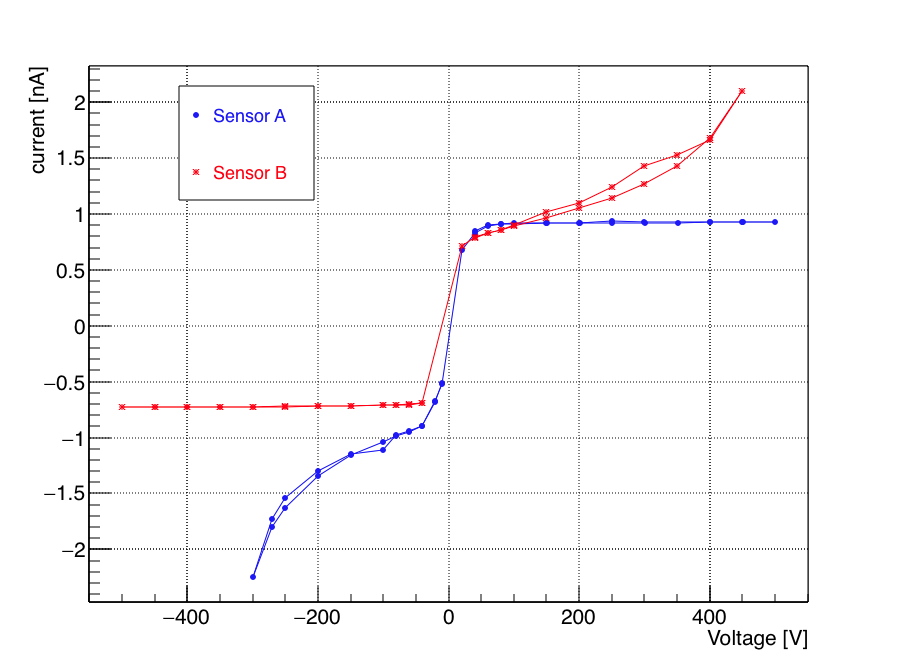}
        \caption{I-V characteristic of two sensors under constant irradiation from a $\beta$ source in a fixed position.}
        \label{fig:iv}
    \end{minipage}
    \hspace{7mm}
    \begin{minipage}{0.45\textwidth}
        \includegraphics[width=2.3in]{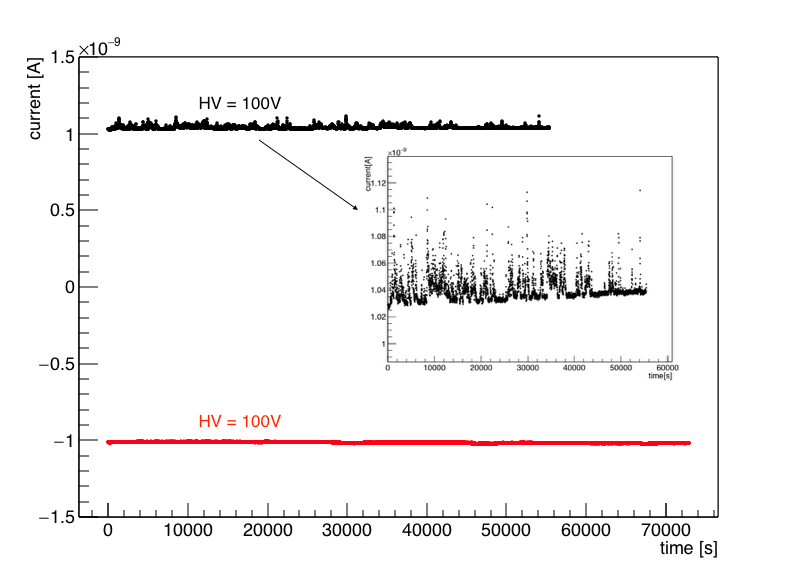}
        \caption{Long term stability measurement for a diamond sensor in two different configurations.}
        \label{fig:stability}
    \end{minipage}
\end{figure}

All devices are pre-tested in terms of dark I-V characteristic. At the typical operation voltage (100V) the dark currents are well below the pA range. Then the measurement of the I-V characteristic (fig.~\ref{fig:iv}) is repeated with sensors irradiated with a $^{90}$Sr $\beta$ source placed at 2mm distance from the sensors, at which we observe a current in the range of about 1nA, that we expect during the operation of the experiment in stable condition of the beams. The measurements are performed biasing the detectors on one electrode and reading the signal on the other. Both polarities are tested. The typical I-V curve observed has a very asymmetric trend with a plateau at one polarity and an increasing current with the opposite bias voltage. Moreover the response of the diamond device has temporary stabilisation issues when the applied voltage is increased or decreased. A hysteresis loop appearing sometime in the I-V curve (fig.\ref{fig:iv}) can be attributed to deep trap levels in the bulk that can act as capture or emission centres of carriers during the charging or discharging process.

As diamond sensors will operate for the entire life of the Belle II experiment long-term stability, under constant irradiation, is checked for periods from several hours to a few days. In fig.~\ref{fig:stability} the response of a diamond sensor as function of time is shown for two different configurations: positive 100V voltage applied at the back and negative 100V voltage applied at the same electrode. In in this example an asymmetric behaviour is observed: 10\% fluctuations for positive voltage, stability within less than 1\% for negative voltage.

\vspace{-0.5cm}
\vspace{-0.2cm}

The current to dose-rate calibration is performed changing the relative distance between the sensor and the source. As for the IV measurement, a dependence of the response on the voltage bias applied is observed: we attribute this to the injection of charge from the electrode to the bulk. In order to relate the current response of the diamond sensor to the dose rate we need to evaluate this ``photoconductive gain''. 
The gain is evaluated as the ratio between measured currents and the current predicted by a simulation by FLUKA~\cite{ref:fluka} that is always smaller than the measured current.

By using the Transient Current Technique (TCT) we can check the crystal quality and the uniformity of the electric field inside the bulk. The $\alpha$ particles from $^{241}$Am radioactive source are used in order to have a localised energy deposit within a few microns, just below the irradiated electrode. According to the bias polarity, one type of the generated charge carriers is quickly collected by the closest electrode, while the other carrier type drifts through the diamond bulk in the direction of the opposite electrode, within a few nanoseconds. The signal observed is then essentially due to only one charge carrier and the properties of electrons and holes are separately measured. For a uniform field in the diamond, in absence of space charge due to carrier trapping, the amplitude of the signal remains constant producing a rectangular current pulse. Deviations from square pulse indicate the presence of a space charge inside the bulk. 

\section{First Phase of SuperKEKB Commissioning}
\begin{figure}[htbp]
    \begin{minipage}{0.45\textwidth}
        \includegraphics[width=2.2in]{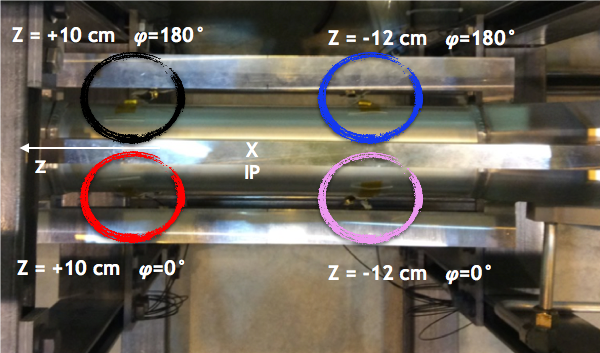}
        \caption{The setup with four diamond sensors installed on the beam pipe for the first commissioning phase.}
        \label{fig:beastSetup}
    \end{minipage}
    \hspace{7mm}
    \begin{minipage}{0.45\textwidth}
        \includegraphics[width=2.3in]{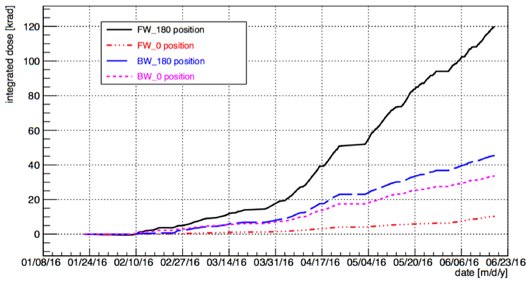}
        \caption{The integrated doses measured in the four positions on the beam pipe during BEAST II phase 1.}
        \label{fig:integratedDose}
    \end{minipage}
\end{figure}

A first commissioning phase (BEAST II phase 1) has been completed in June 2016. A prototype detector system was installed in order to provide measurements of the background levels, whose predictions by simulation have large uncertainties. During phase 1 four diamonds (fig.~\ref{fig:beastSetup}) and the prototype readout electronics were in place and operational. The first results of the analysis of data from this first phase show a clear correlation with different beam conditions. The diamond sensors were very sensitive to the beam losses correlated with current value, the number of bunches and beam size. Their stability and sensibility will allow to implement beam abort features in Phase II. A first measurement of the integrated doses relative to the four diamond positions is shown in fig.~\ref{fig:integratedDose}. 

\vspace{-0.3cm}

\section{Conclusions}
The SuperKEKB collider will provide unprecedented high luminosity. As a consequence severe beam-induced backgrounds and radiation doses are expected. A radiation monitoring and beam abort system based on twenty single-crystal diamond sensors will be adopted in the VXD volume. As the response of diamond sensors as dosimeters can be influenced by several issues a complete characterisation is performed in order to relate the measured current with the dose rate.

%
%

\clearpage
\addtocmark[2]{Author Index} 
\renewcommand{\indexname}{Author Index}
\printindex
\clearpage
\end{document}